\documentclass{appolb}
\usepackage{bbm}
\usepackage{psfrag}
\usepackage{graphicx}
\begin{document}

\title{Fermions in 2D Lorentzian Quantum Gravity}

\author{L. Bogacz, Z. Burda and J. Jurkiewicz
\address{M. Smoluchowski Institute of Physics, Jagiellonian University,
Reymonta 4, 30-059~Krak\'ow, Poland. }
}

\maketitle

\begin{abstract}
We implement Wilson fermions on 2D Lorentzian triangulation
and determine the spectrum of the Dirac-Wilson operator. We compare it
to the spectrum of the corresponding operator
in the Euclidean background. We use fermionic particle
to probe the fractal properties
of Lorentzian gravity coupled to $c=1/2$ and $c=4$ matter.
We numerically determine the scaling exponent
of the mass gap $M \sim N^{-1/d_H}$ to be $d_H=2.11(5)$, and 
$d_H=1.77(3)$
for $c=1/2$ and $c=4$, respectively.
\end{abstract}

\section{Introduction}

The formulation of a theory of quantum gravity
is one of the most challenging problems in theoretical physics.
Simplicial gravity is a non-perturbative
approach to this problem.
It is a natural extension of Feynman's idea of
defining quantum amplitudes via functional path integrals.

Simplicial gravity is a lattice regularization
of Feynman integrals over a set of geometries \cite{aj,am,b}.
The idea is to look for non-perturbative fixed points
of the renormalization group at which a
continuum limit can be taken. 
It is crucial in a lattice regularization to preserve
the gauge invariance of the underlying continuum theory.
Simplicial gravity, similarly as lattice
regularization of QCD, properly treats the problem
of gauge invariance. The underlying continuum theory
is invariant with respect to the change of coordinates.
Lattice formulation is coordinate free by construction. 
A remnant of coordinates are labels
on lattice simplices and vertices. Lattice theory is defined
in a way which is invariant with respect to relabeling.
This invariance is a left over of the diffeomorphism  invariance
of the continuous formulation. Statistical weights of
simplicial manifolds take into account
the volume of the discrete symmetry group. For
example, in two dimensions, were the sum over simplicial
diagrams (dynamical triangulations) can
be explicitly generated by a perturbative expansion of a
matrix model \cite{bipz,d}, the statistical weights are automatically
generated as combinatorial factors of the corresponding
Feynman diagrams. These factors play the role of the
Faddeev-Popov determinants. The 2d case is
analytically solvable. The continuum limit
of 2d lattice gravity \cite{bk} is equivalent
to Liouville theory being a quantum version of Euclidean
gravity regularized by  completely different
means \cite{p,kpz}.
This equivalence is treated as a strong indication that the sum
over simplicial manifolds provides a proper definition of
the integration measure over Riemannian manifolds.

Real gravity has the Lorentzian signature. One can obtain
this signature by Wick rotation. One way of doing this
is to calculate quantities in the Euclidean sector and then
perform analytic continuation to the Lorentzian one.
This strategy is used in quantum field theory but it is
not clear whether it can be straightforwardly
applied to quantum gravity. An alternative approach 
is to impose the causal structure on simplicial 
manifolds which enter the Feynman integrals \cite{al,ajl}. 
This leads to a formulation called Lorentzian simplicial gravity 
for which the causality is achieved by introducing 
a time-slicing into the lattice structure. This formulation is 
very close in spirit to the Hamilitonian formulation in 
the temporal gauge.
In two dimensional case one can determine an explicit form of
the Hamiltonian of the underlying continuum theory \cite{dl}.
Similarly as for the Euclidean case the model 
is analytically solvable in two dimensions \cite{dl,fgk,ackl}.
The resulting continuum theory differs from Liouville
gravity. One can determine mathematical relations between
Euclidean and Lorentzian gravity in terms of 
a singular renormalization of coupling constants \cite{ackl}.

Both the Euclidean \cite{aj,am,b} and
Lorentzian models \cite{ajl1,ajl2,ajlv}
have natural extensions to higher dimensional cases.
It is a matter of debate which of the two versions may
serve as a theory of quantum gravity in higher dimensional
case. Both have been a subject of
intensive studies. In the ultimate theory of
gravity an important role is played by the interaction
of gravity with matter fields. Matter fields are known to modify
fractal properties of gravity
and scaling properties of the underlying theory.
Results of explicit calculations of 2d Euclidean case,
are summarized in the
KPZ formula \cite{kpz}. They show that the scaling
properties of matter field are indeed modified by Euclidean
gravity. On the contrary, numerical simulations of Lorentzian
case indicate that the scaling properties of matter in
Lorentzian background stay intact
even if one crosses $c=1$ barrier \cite{aal1,aal2}.

Numerical studies of higher-dimensional Euclidean gravity
have shown the importance of matter fields for 
of the critical properties of the underlying continuum theory.
For example, matter fields
remove the conformal instability of Euclidean sector
and modify the phase structure of the model \cite{bbkp,bbkptt1,bbkptt2}. 
So far numerical simulations
have been performed only for bosonic matter. An introduction of
fermionic matter may be crucial for defining a fixed point of gravity
at which a continuum limit can be taken.
First step towards defining fermions on
simplicial quantum gravity was done in 2d Euclidean gravity 
\cite{bjk,bbjkpp,bbpp}.
In this paper we extend these studies to the 2d Lorenzian case.
We shall use fermions to probe fractal properties of geometry.

\section{The model}

Let us briefly recall the model of 2d Lorentzian gravity \cite{al,ajl}.
The integration measure of the 2d Lorentzian gravity is
defined as a sum over triangulations which have a 
time sliced structure.
Additionally, for technical reasons this structure is periodic
in temporal direction. 

Each time slice consists of a random number 
of vertices on a circle. The
number of vertices $V_{t}$ on a slice $t$ and $V_{t+1}$
on the consecutive slice
give the number of triangles $N_{t+1/2} = V_t + V_{t+1}$ lying in between.
The strip between slices consists of a random 
combination of triangles built of
edges which join vertices of the two time slices.
The temporal index runs periodically over $t=1,..,T$.
Consequently, the total numbers of triangles and
vertices are related to each other as $N=2V$. 
Topology of each time slice
is that of circle in contrast to Euclidean gravity
where it can be a set of disconnected circles.
The effect of branching, which plays a dominant role in
Euclidean case, is thus suppressed here.
In consequence, fractal properties of Lorentzian gravity
are completely different from those of Euclidean gravity,
as reflected by the Hausdorff dimension which changes
from $d_H=2$ \cite{aal1,aal2} for the former to
$d_H=4$ \cite{kkmw} for the latter case.
In a sense, Lorentzian gravity has not enough
freedom to produce structures which would significantly
deviate from flat geometry of the canonical dimension $d=2$.
It is generally very difficult to change fractal properties
of the Lorentzian gravity. This can be achieved by strengthening
the influence of the matter sector on geometry by increasing
its conformal charge $c$. It was shown in \cite{aal1,aal2}
that a multiple Ising field with
$q$-families, and the conformal charge $c=q/2>1$,
modifies fractal properties of the underlying geometry leading to
a space-time with the Hausdorff dimension $d_H\approx 3$. The MC
simulations \cite{aal1,aal2} were done for $c=4$ which is
for technical reasons an optimal choice~: $c=4$
is large enough to allow for observing 
for relatively small lattices the effects
of crossing the $c=1$ barrier 
and on
the other hand it is still not very large from the point of 
view of MC simulations, in particular of the computer 
time needed to update the matter sector, which
grows linearly with $c$. We stick here to $c=4$.

As mentioned we shall use fermionic
particle to probe geometrical properties of the Lorentzian
background. More precisely, we
shall do this by studying the scaling properties of the lowest part
of the spectrum of the Dirac-Wilson operator.

We consider a system of $q$ species of Ising fields on dynamical
Lorentzian triangulations with $N$ triangles and $T$
time slices. The  canonical partition function of this system reads~:
\begin{equation}
Z_{(q)}(\beta) = \sum_{l \in L_{N,T}} \bigg( Z_l(\beta) \bigg)^q
\label{Z}
\end{equation}
where the sum runs over all triangulations $l$
from the set of Lorentzian triangulations $L_{N,T}$ with $N$ triangles
and $T$ time slices. Each triangulation $l$ is dressed with
$q$ species of independent Ising spins and thus the weight 
of each triangulation in the ensemble is given by the
$q$-th power of the partition function of a 
single Ising field on this triangulation~:
\begin{equation}
Z_l(\beta) = \sum_{\{\sigma\}_l} 
\exp \big(\beta \sum_{(ab)} \sigma_a \sigma_b \big)
\end{equation}
Here $a,b$ and $(ab)$ denote vertices and links  of the Lorentzian
triangulation $l$, respectively. Spins live on vertices.
Each spin $\sigma_a$ assumes two values $\sigma_a = \pm 1$.
The sum $\{\sigma\}_l$ runs over all $2^{N/2}$ spin 
configurations of one spin family on the lattice $l$.
Although spin families are independent
on a given triangulation, 
they are not independent in the ensemble of triangulatons
since they interact through dynamical lattices,
which are summed over in the partition function (\ref{Z}).

The partition function for an individual spin family can be
rewritten as a partition function for Ising spins living
on vertices of the dual lattice $l$ or equivalently
on triangles of the original lattice.
The dual temperature $\beta_*$
is related to $\beta$ as  $\tanh \beta_* = \exp - 2\beta$.
The equivalence between the original and dual model holds
up to finite size effects \cite{bbjkpp}.
The Ising model is also equivalent to a model of Wilson
fermions for Majorana fields located on triangles~:
\begin{equation}
{\cal Z}(K) = \sum_{l \in L_{N,T}} \bigg( {\cal Z}_l(K) \bigg)^q
\end{equation}
where each ${\cal Z}_l$ stands for
for the partition function for Majorana fermions
on a lattice $l$~:
\begin{eqnarray}
{\cal Z}_l(K) = 
\int \prod_{i} d \bar{\Psi}_{i} d\Psi_{i}
\exp \big( - \frac{1}{2} \sum_{i} \bar{\Psi}_i \Psi_i +
K \sum_{\langle ij \rangle} \bar{\Psi}_i H_{ij} \Psi_j \big) \\ \nonumber 
= \int \prod_{i} d \bar{\Psi}_{i} d\Psi_{i}
\exp \big( - \sum_{ij} \bar{\Psi}_i D_{ij} \Psi_i\big) 
\label{ZF}
\end{eqnarray}
with fermions located at the centers of triangles $i,j,\dots$.
The Dirac-Wilson operator $D = \mathbbm{1}/2 + K H$ consists
of a mass part and a hopping term controlled by the hopping
parameter $K$. The sum in the hopping term runs over all oriented pairs
$\langle ij \rangle$ of nearest triangles on the triangulation $l$.
The hopping operator $H_{ij}$
can be expressed in local frames
as~:
\begin{equation}
H_{ij} = \frac{1}{2} \left( 1 + \vec{n}_{ij} \vec{\gamma} \right)
{\cal U}_{ij}
\end{equation}
where $\vec{n}_{ij}$ is a vector of the local derivative
which goes between the neighbours $i$ and $j$
and ${\cal U}_{ij}$ is a spin connection
in the spinorial representation. 
The components of the spinors $\Psi_i$ are given in the local frames.
The spin connection matrices allow for parallel transport of spinors
between neigbouring frames and for recalculating spinor components. 
The hopping parameter $K$ is related to the Ising temperature as~:
\begin{equation}
K = \frac{e^{-2\beta}}{\sqrt{3}} = \frac{\tanh(\beta_*) }{\sqrt{3}}
\end{equation}
The critical temperature of the Ising model corresponds to
the critical value of the hopping parameter for which fermions
become massless. The critical value for the Euclidean gravity
can be analytically determined
$\beta_{cr}=\frac{1}{2}\ln \frac{131}{85}=0.21627...$ \cite{bj}.
It corresponds to the critical value of the hopping parameter
$K_{cr} = 85\sqrt{3}/393 = 0.3746....$ which should 
be compared with the critical value on the regular 
triangulated lattice~: $K_{cr}=1/3=0.3333...$ \cite{bj}.
As one can see, the interaction with a random lattice 
dresses the critical value of the hopping parameter 
similarly as interactions with gauge fields for QCD.
As we shall see the dressing of the hopping parameter is different for
Lorentzian gravity.

The matrix of the Dirac-Wilson operator can be easily read off
from the equation (\ref{ZF}).
The spectrum of the Dirac operator is related to the propagation
of a fermionic particle through the lattice. The smallest
eigenvalues are related to the effective mass of this particle.
For an infinite lattice and at the critical
value of the hopping parameter the theory describes
a massless Majorana fermion. For a finite lattice there 
exists a non-vanishing
mass gap which separates the lowest part of the spectrum
from zero. This mass gap is minimal for certain value of
the hopping parameter which we will refer to as pseudo-critical.
We will denote this value as $K_{*}$ and the corresponding mass
gap as $M_*$. The two values change with the lattice size $N$
and are expected to approach their limiting values
$K_* \rightarrow K_{cr}$ and $M_* \rightarrow 0$
for $N \rightarrow \infty$. In particular
one expects the scaling~:
\begin{equation}
M_* \sim N^{-1/d_H}
\label{me}
\end{equation}
with the exponent $d_H$ which is related to the fractal
properties of the underlying geometry. 
For an isotropic system, like for instance Euclidean gravity
on a regular lattice, this exponent corresponds to the Hausdorff 
dimension $D_H$.
Lorentzian lattice is anisotropic. Its fractal dimensions in the temporal
and spatial directions change with the matter content \cite{aal1,aal2}. 
The spatial and temporal asymmetry becomes very transparent 
when one crosses the $c=1$ barrier.
In this case the system forms a bubble which is supplemented by a narrow 
long neck. Denote the temporal extension of the bubble by $T_B$ and spatial 
by $L_B$. The temporal and spatial extensions of the bubble scale differently
with the size, $N_B$, of the bubble $T_B \sim N_B^{1/D_H}$ and 
$L_B \sim N_B^{1/\delta_h}$.
The fractal dimensions $D_H$ and $\delta_h$ are not independent. Using
the relation $N_B \sim T_B L_B$ one can see that
\begin{equation}
\frac{1}{D_H} + \frac{1}{\delta_h} = 1
\label{asym}
\end{equation}
In particular, for $c=1/2$ the 
two exponents are $D_H=2$ and $\delta_h=2$, merely
reflecting the fact that the bubble is not developed and 
the temporal size of the bubble corresponds to the temporal extension of the
system $T_B \sim T$ and correspondingly $L_B \sim N/T$. The situation changes
dramatically for $c=4$. In this case, $D_H=3$ and $\delta_h=3/2$. 
In view of this asymmetry the following question arises. The scaling of the
lowest part of the spectrum of the Dirac operator is expected to be
controlled by the lowest momentum and thus in this case one can expect
the mass exponent to be $d_H=\delta_h$.
On the other hand as discussed in \cite{aal1,aal2}
matter fields coupled to Lorentzian gravity even above $c=1$ barrier have
flat space exponents which means that the fields behave effectively as in 
a flat 2d background. According to this hypothesis one should observe the
value $d_H=2$ of the mass exponent. Which of the scenarios is 
realized in the system, is one of the questions addressed here.

\section{Numerical set-up}

Let us shortly describe our 'experimental' set-up.
We use a MC generator to simulate a system of a
given size $N$ and a given temporal extent $T$.
The average number of vertices per slice is $L=N/2T$ and hence the
lattice asymmetry is $\tau = T/L = 2T^2/N$. The bulk thermodynamic
properties of the system are expected to be independent of $\tau$.
This parameter can be thus used to minimize finite size effects.

Geometry of the lattice is updated by the standard local
algorithm based on a pair of mutually reciprocal moves~:
split and join operations \cite{aal1}. The transformations preserve
the temporal length of the system $T$ but change the lattice size $N
\leftrightarrow N+2$. In order to ensure ergodicity of this algorithm
one allows the system size to fluctuate.
In practice one does it by simulating a system with
a partition function
\begin{equation}
z(\beta) = \sum_{l \in L_T}
e^{-\lambda n -
\frac{1}{2\sigma^2} (n - N)^2  }  \bigg( Z_l(\beta)\bigg)^q
\label{tZ}
\end{equation}
with a volume $n$ which may fluctuate. In order to avoid
too large fluctuations 
an external potential $U(n) = \lambda n + (n-N)^2/(2\sigma^2)$
is added to the action in (\ref{tZ}). This potential
constrains the volume fluctuations to a neighbourhood of $N$.
The width of the distribution of $n$ is of order $\sigma$.
If the parameter $\lambda$ is optimally tuned, the maximum of the
distribution lies exactly at $N$. 

The algorithm generates a smeared distribution of
volumes $n$ but measurements are performed only
at $n = N$. The condition $n=N$ cuts out from the ensemble (\ref{tZ})
a sub-ensemble with the partition function which is equal to
(\ref{Z}) up to a constant factor irrelevant for statistical 
averages at $N$.

We concentrate the MC measurements on the spectrum of the
Dirac-Wilson operator $D_{ij}$ (\ref{ZF}). 
Each triangle on the lattice is dressed with
a two-component spinor, and hence for a lattice with $N$ 
triangles the operator is represented by a
$2N\times 2N$ matrix.
The evaluation of the spectrum requires a time
proportional to $N^3$. This is a very time consuming operation.
We use the reduction to the Hessenberg form
and then QR decomposition procedure \cite{gvl}.
Many problems of interest are related to the behaviour of
the lowest part of the spectrum. We use the Lanczos algorithm
to determine the position of the lowest eigenvalues.
The Lanczos algorithm is most efficient for this purpose \cite{gvl}.

\section{Results}

The spectrum of the Dirac-Wilson operator is complex.
As far as the lowest part of the spectrum is concerned it is
more convenient to study the Majorana-Wilson operator
instead of the Dirac-Wilson one
\begin{equation}
{\cal D} = C D
\label{cD}
\end{equation}
because it has a purely imaginary spectrum. $C$ is the charge conjugation
matrix. Indeed, if one chooses a representation
in which the two-dimensional  $\gamma$ matrices are real~:
$\gamma_1=\sigma_3$, $\gamma_2=\sigma_1$,
so is the charge conjugation matrix,
$C= i\sigma_2 = \epsilon$ and the whole matrix of the 
Majorana-Wilson operator ${\cal D}$. 
Since the matrix ${\cal D}$ is also antisymmetric, 
it is anti-Hermitian.
>From here on, when we refer to the lowest eigenvalues,
we mean the closest to zero eigenvalues of the Majorana-Wilson operator.
In fact, as mentioned already, it is rather this operator than the
Dirac-Wilson one, which is related to the Ising spins and the
conformal field with $c=1/2$.

The construction of the Dirac-Wilson operator requires an introduction
of a field of local frames on a simplicial manifold which locally
defines gamma matrices and a spin connection ${\cal U}$.
An explicit construction of the operator is given in \cite{bjk,bbjkpp}.

\begin{figure}
\begin{center}
\psfrag{xx}{Re$\lambda$}
\psfrag{yy}{Im$\lambda$}
\includegraphics[height=5.5cm]{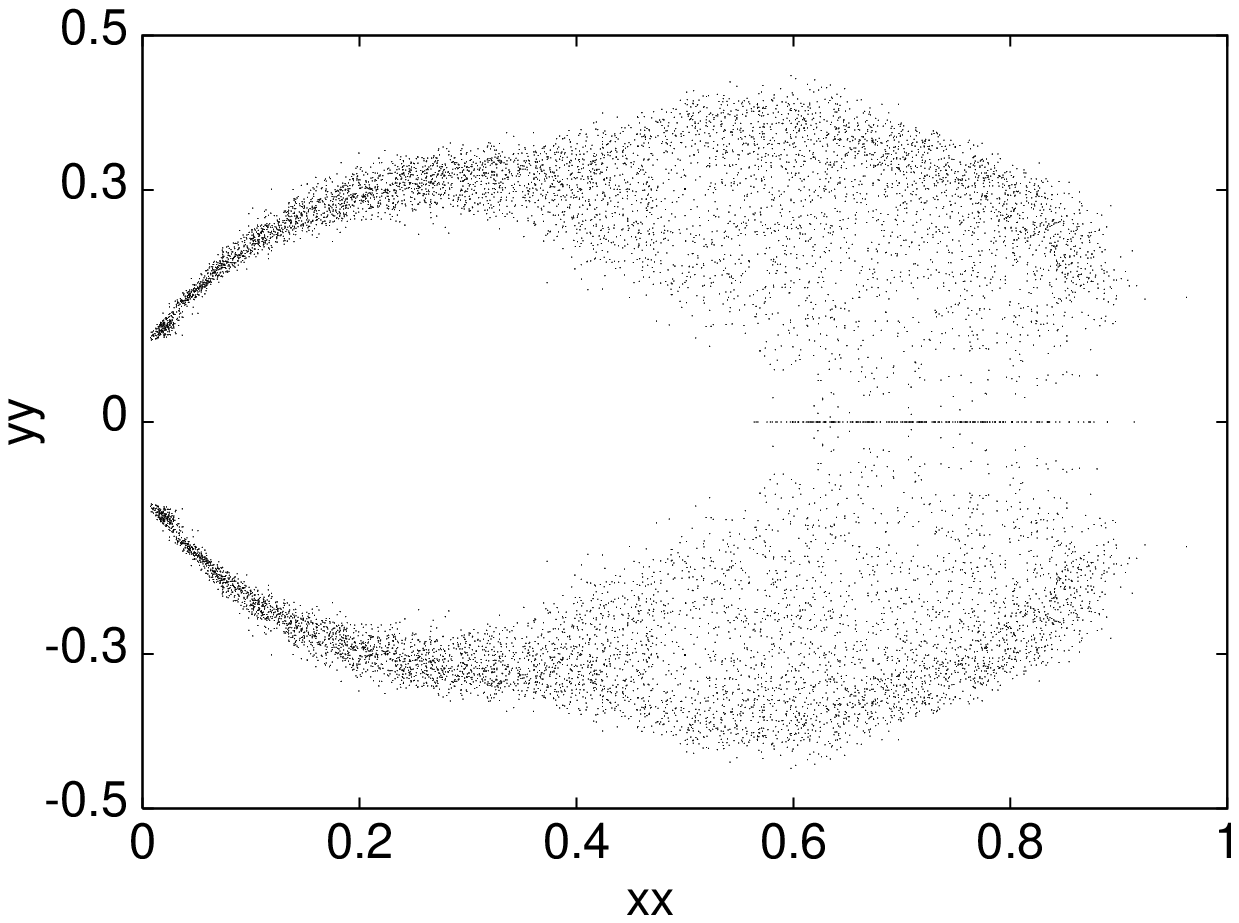}\\
\vspace{0.2cm}
\includegraphics[height=5.5cm]{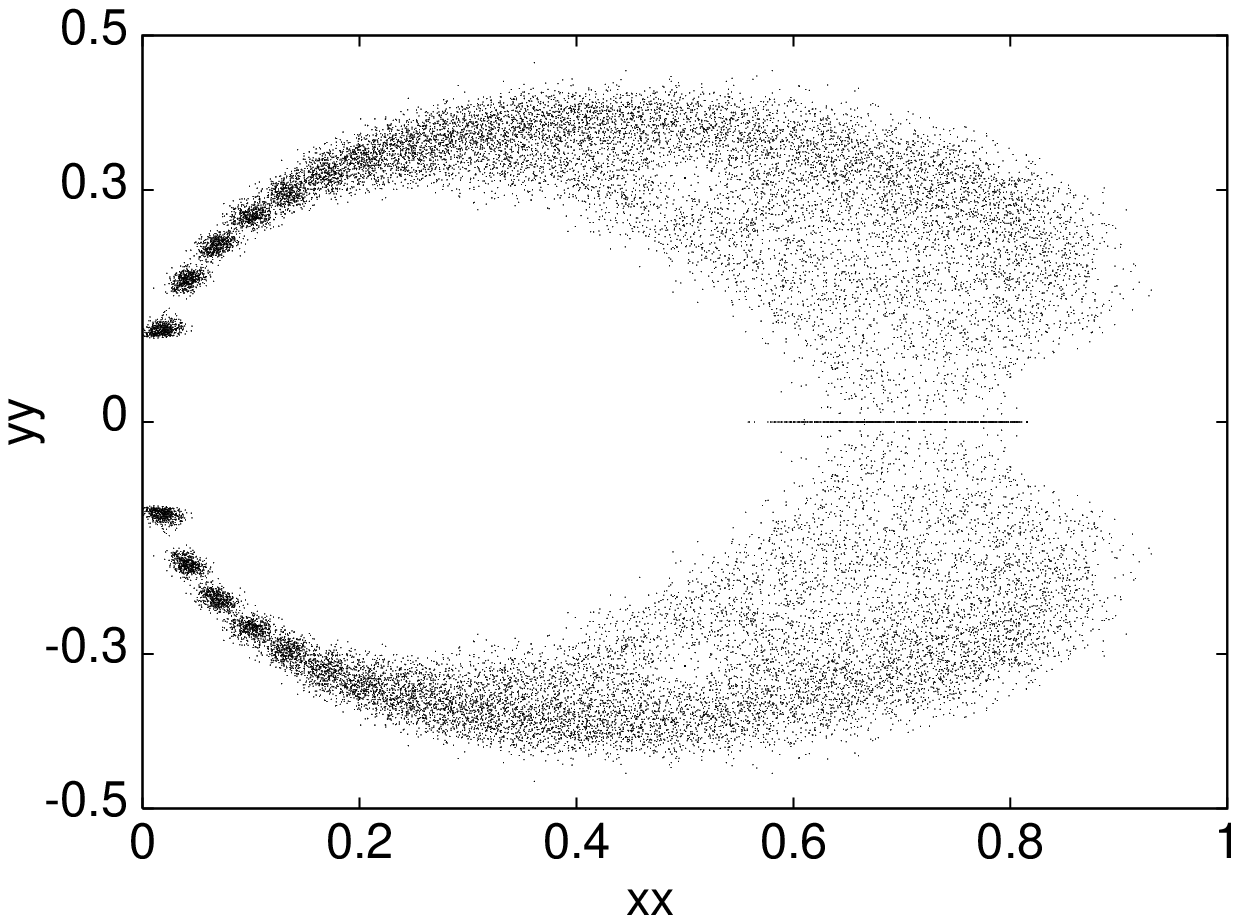}
\caption{\label{spectrumL} Spectra of
the Dirac-Wilson operator measured in MC simulations
for Lorentzian lattice with $N=64$ triangles,
$\tau=2$ and $K=0.3486$ (top),
and for Euclidean lattice with $N=64$ triangles,
$K=0.364$ (bottom).}
\end{center}
\end{figure}

A snapshot of the spectrum of the Dirac-Wilson operator generated in a MC
simulation of Lorentzian gravity is shown in
fig.\ref{spectrumL}. It should be compared with the spectrum
on Euclidean lattice. As one can see there are some
visual differences for eigenvalues with large absolute values. 
The differences obviously have the origin
in the different properties of Lorentzian and Euclidean lattices 
on small distances. What is of physical interest is 
the small eigenvalue behaviour of the spectrum because it is responsible 
for the large distance behaviour
and the universal critical properties of the system. This behaviour
is governed by the scaling of the part of the spectrum closest 
to the origin of the complex plane. 
As will be shown, it is given by the mass exponent which has a different
value for the Lorentzian than for the Euclidean case.
The spectrum changes with the hopping parameter $K$.
The main effect of this change on the shape of the
spectrum is that it gets rescaled in the complex plane
around the point $(1/2,0)$ as follows directly
from the form of the operator $\mathbbm{1}/2 + K H$
which is a sum of a constant operator $\mathbbm{1}/2$ and a
random operator $H$ multiplied by the factor $K$.
In addition to this effect, the matter sector influences 
geometry of the lattice and hence also the randomness encoded 
in the operator $H$. 

\begin{figure}
\begin{center}
\psfrag{yy}{$M$}
\psfrag{xx}{$K$}
\includegraphics[height=5cm]{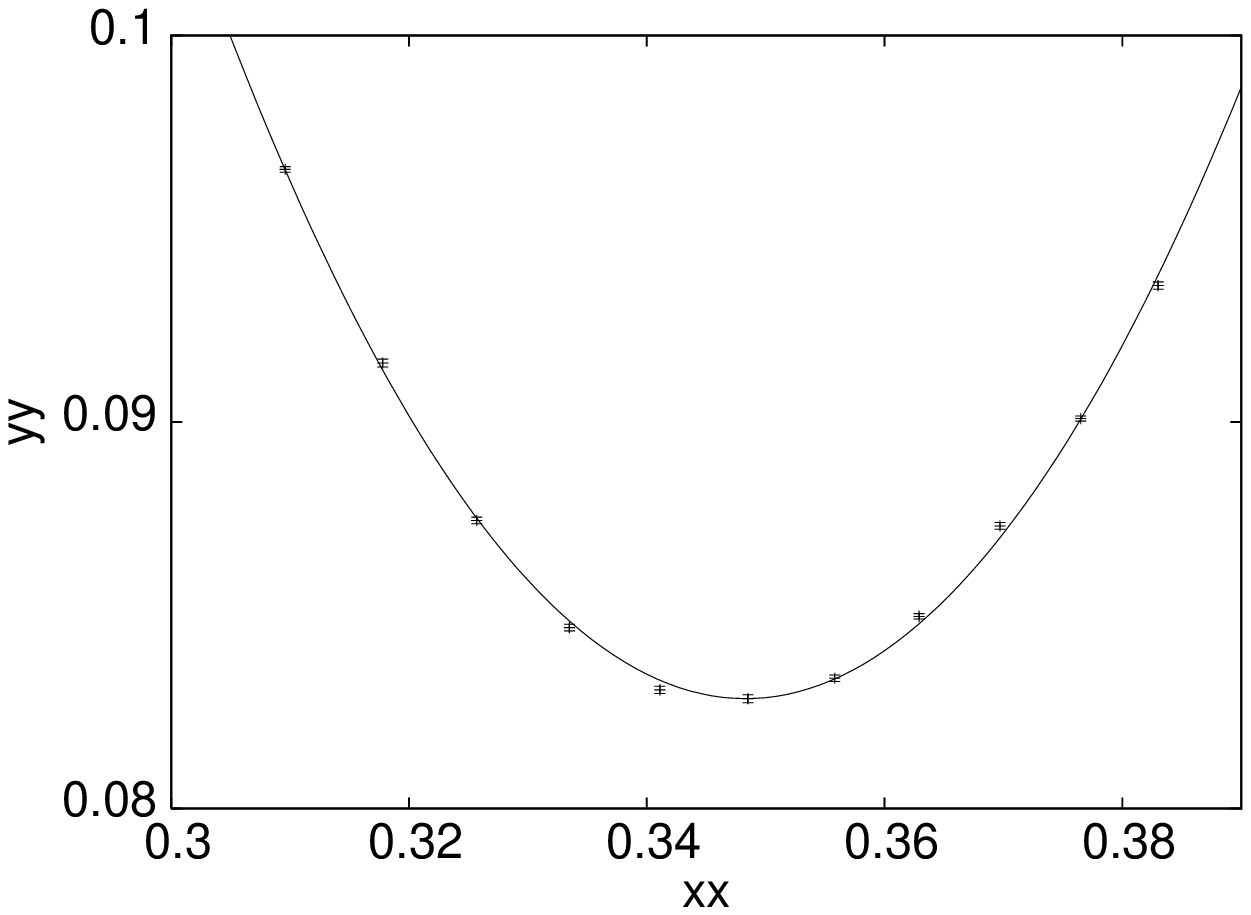}
\caption{\label{M} Mas gap $M$ as a function 
of hopping parameter $K$ for a lattice with $N=200$ triangles,
asymmetry parameter $\tau=4$ and the conformal charge $c=1/2$.
A parabola is fitted to the data.}  
\end{center}
\end{figure}

When the hopping parameter changes from $K=0$
(which corresponds to $\beta=\infty$) to $K=1/\sqrt{3}$
($\beta=0$) the spectrum broadens from a spectrum
localized at the point $(1/2,0)$
to an extended shape. In the course of the broadening
the claws-shaped part of the spectrum passes close to the
origin of the complex plane.  The value of $K$, at which
the distance of eigenvalues to the origin
is smallest, corresponds to a pseudo-critical
value $K_*$. A similar effect is seen in the movement of
the lowest end of the spectrum of the Majorana-Wilson
operator which first moves towards zero when $K$ grows
from zero to $K_*$ and then moves away from zero when $K_*$
further increases. We use this observation to determine the
mass gap $M_*$ for the system with a given volume in
the following way. We determine
the mean-value $M$ of the distribution of the lowest
eigenvalue of the Majorana-Wilson operator for the system
with a size $N$ and a hopping parameter $K$. Then we plot the
dependence
$M=M(N,K)$ as a function of $K$ (see fig.\ref{M}).
Combining the quadratic interpolation with the jack-knife
method  we find the minimum $M_*(N)$ of the plotted function
We repeat the same procedure for different volumes $N$ to obtain 
the dependence of the pseudo-mass on the volume of the system $M_*(N)$.
Eventually we fit the experimentally determined
points $M_*(N)$ to the
scaling formula (\ref{me}) to determine the optimal value of
the scaling exponent $d_H$.  
\begin{figure}
\begin{center}
\psfrag{yy}{$M_*$}
\psfrag{xx}{$N$}
\includegraphics[height=5cm]{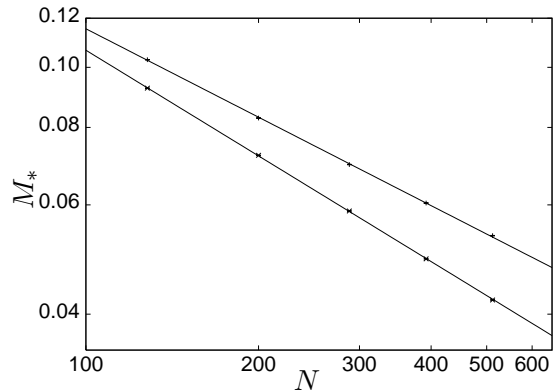}
\caption{\label{M*} Mas gap $M_*$ of fermionic particle
as a function of lattice volume $N$, 
for lattices with deformation parameter $\tau=4$ and conformal charge 
$c=1/2$ (upper line) and $c=4$ (lower line). 
The fit $M_*(N)= a N^{-1/d_H}$ gives $a=1.02(2)$ 
and $d_H=2.11(5)$ for $c=1/2$ and  $a=1.44(2)$ and $d_H=1.77(3)$}.  
\end{center}
\end{figure}
As an example, in fig.\ref{M*} we show the experimental data points
and the best fits to (\ref{me}) for 
$c=1/2$ and $c=4$.
The system size varies in the range from $N=128$ to $N=512$ and
the asymmetry parameter is $\tau=4$ for the presented data.
A value of the pseudo-critical hopping parameter depends on the
matter content. For example for the lattice size $N=128$ it is 
$K=0.3482(2)$ and $K=0.3536(8)$ for $c=1/2$ and $c=4$ respectively. 
The corresponding value for Euclidean 
lattice of the same size is $K=0.370(1)$.

The shape of the spectrum of the Dirac-Wilson operator
for the pseudo-critical value of the hopping parameter $K$ 
does not change visually when the system size $N$ grows,
except for the two claw ends
which approach the origin of the complex plane as $N^{-1/d_H}$. 
The two ends eventually close at the origin
when $N$ becomes infinite. This effect corresponds to the
appearance of a massless particle on an infinite lattice. 
The shape of the bulk part of the spectrum does not change but merely
becomes denser for larger lattices, which means that the typical
distance between eigenvalues becomes smaller for larger system volumes.

The best fit to the scaling formula (\ref{me}) gives a value
$d_H=2.11(5)$ for $c=1/2$. This value should be compared with 
$d_H=2.87(3)$ measured for Euclidean gravity interacting
with the $c=1/2$ matter \cite{bbpp}. 
Fermionic particle in the Lorentzian background,
in contrast to Eucildean gravity, detects thus a 
flat space exponent.

For the case $c=4$ for which geometry of the lattice  
changes dramatically, the value of the mass exponent is $d_H=1.77(3)$.
Its value clearly moves towards the spatial scaling dimension
$\delta_h=3/2$ which is dictated by asymmetry of  
fractal properties of the lattice (\ref{asym}). 
As mentioned before, Lorentzian lattice coupled with matter field
for $c=4$ consists of two distinct parts~: of the
bubble whose extensions scale as $T_B \sim V_B^{1/D_H}$ and
$L_B \sim V_B^{1/\delta_h}$ and of a narrow neck whose spatial
width does not scale.
The presence of the narrow neck introduces a finite size effect
to the measurements of the spectrum of the Dirac operator.
This effect is probably the source of the deviation of the measured
value $d_H=1.77(3)$ from $\delta_h=3/2$. 
One should try to reduce the finite size finite size effect by going to
larger lattice for which the contribution of the neck
to the spectrum should gradually decrease because 
the number of triangles on the bubble grows much faster
than the number of triangles in the neck.
This is however very
time consuming because the time for collecting the spectrum grows
typically as the third power of the system size.

\section{Summary}
We implemented fermions to Lorentzian gravity
and determined the spectrum of Dirac-Wilson operator.
We calculated the mass gap exponent 
$d_H=2.11(5)$ for a single massless fermionic particle 
propagating in the Lorentzian background and interacting with it.
The computations were done for lattices of size up to $N=512$ 
triangles.
The measured value of the mass exponent seems to be
consistent with the canonical dimension $d=2$, if one takes into
account a possibility of finite size effects. For $c=4$,
above the $c=1$ barrier, we measured $d_H=1.77(3)$. The value of the
exponent moved towards the index $\delta_h=3/2$ which controls the
scaling of the spatial momentum.
In fact, if we explicitly introduce a finite size correction
to the formula (\ref{me}) of the form $M_* \sim N^{-1/d_H}(1 + c/N)$
the value of the exponent gets shifted to $d_H=1.7(1)$ indicating 
indeed the presence of a deviation from the straight line 
in the measured volume range. It would be very helpful
to extend the simulations 
to larger systems to see whether the observed tendency will indeed
bring the exponent to the expected value $3/2$. As we mentioned, 
the measurements of the spectrum for larger volume are very 
time consuming due to the strong dependence of the required CPU 
time on the volume. We plan to perform these computation in the
future. 

The measurements of the spectrum of the Dirac operator 
enable one to directly detect in the matter sector 
changes of the fractal structure of two-dimensional Lorentzian gravity. 
In other measurements of the critical indices of the matter
sector one namely  sees the flat space critical exponents 
even above the $c=1$ barrier,
for example, the Onsager exponents \cite{aal1,aal2}
for the Ising model. 
The spectrum of the Dirac-Wilson operator is sensitive to 
changes of fractal structure and therefore it provides 
a practical tool for a detection of fractal properties 
of the geometrical background.

\section{Acknowledgments}

This work was partially supported by Polish State Committee for
Scientific Research (KBN) grants: 2P03B 09622 (2002-2004),
2P03B 00624 (2003) and by EU Network HPRN-CT-1999-00161.

\end{document}